\documentclass{llncs}
\usepackage{amsmath}
\usepackage{makeidx}  % allows for indexgeneration

\usepackage{times}
\usepackage{latexsym}
\usepackage{url}
\usepackage{graphicx}
\usepackage{flexisym}
\graphicspath{ {images/} }
\usepackage[colorinlistoftodos]{todonotes}
\usepackage{float} 
\usepackage{multirow}
\usepackage{aurl}
\usepackage{cite}
\usepackage{cleveref}

\usepackage{makeidx}  % allows for indexgeneration
\usepackage{times}
\usepackage{url}
\usepackage{latexsym}
\usepackage{url}
\usepackage[utf8]{inputenc}
\usepackage[T1]{fontenc}
\usepackage{graphicx}
\usepackage{listings}
\usepackage{color}
\usepackage[usenames,dvipsnames,svgnames,table]{xcolors}
\usepackage{amsfonts}
\usepackage{mathdots}
\usepackage{float}
\usepackage{framed}
\usepackage{todonotes}
\usepackage{wrapfig}
\usepackage{tikz}
\usetikzlibrary{positioning}
\usepackage{comment}
\usepackage{times}
\usepackage{wrapfig}
\usepackage{lscape}
\usepackage{rotating}
\usepackage{epstopdf}
\usepackage{amssymb} 
\usepackage{pgfplots}
\usepackage{subcaption}
\usepackage{verbatim}
\usepackage{booktabs}
\usepackage{tabularx,ragged2e,booktabs}
\newcolumntype{L}{>{\RaggedRight\arraybackslash}X} % ragged-right version of "X"
\usepackage{pgfplots}
\usepackage{enumitem}
\usepackage{epstopdf}
\usepackage{CJKutf8}
\usepackage[ruled,vlined,linesnumbered]{algorithm2e}

\pgfplotsset{compat=1.14}

\begin{document}
\pagestyle{headings}  % switches on printing of running heads
	
\title{Named Entity Recognition in Twitter using Images and Text}
\titlerunning{Named Entity Recognition in Twitter}

	%\runningtitle{Instructions for the preparation of a camera-ready paper in \LaTeX}
	%\subtitle{Subtitle}
	
	\author{Diego Esteves\inst{1} \and Rafael Peres\inst{2} \and Jens Lehmann\inst{1,3} \and Giulio Napolitano\inst{1,3}}
	\authorrunning{Diego Esteves et al.} % abbreviated author list (for running head)
	\institute{University of Bonn, Germany\\
	\email{\{surname\}@cs.uni-bonn.de},\\
	\and
	Federal University of Rio de Janeiro, Brazil\\
	\email{rafaelperes@ufrj.br}
	\and
	Fraunhofer IAIS, Bonn, Germany\\
	\email{giulio.napolitano@iais.fraunhofer.de}}
	\maketitle              
	
	\begin{abstract}
	Named Entity Recognition (NER) is an important subtask of information extraction that seeks to locate and recognise named entities. Despite recent achievements, we still face limitations with correctly detecting and classifying entities, prominently in short and noisy text, such as Twitter. 
	An important negative aspect in most of NER approaches is the high dependency on hand-crafted features and domain-specific knowledge, necessary to achieve state-of-the-art results. Thus, devising models to deal with such linguistically complex contexts is still challenging. In this paper, we propose a novel multi-level architecture that does not rely on any specific linguistic resource or encoded rule. Unlike traditional approaches, we use features extracted from images and text to classify named entities. Experimental tests against state-of-the-art NER for Twitter on the \textit{Ritter} dataset present competitive results (0.59 F-measure), indicating that this approach may lead towards better NER models.
		\keywords{NER, Short Texts, Noisy Data, Machine Learning, Computer Vision}
	\end{abstract}

\section{Introduction}

Named Entity Recognition (NER) is an important step in most of the natural language processing (NLP) pipelines. It is designed to robustly handle proper names, which is essential for many applications. Although a seemingly simple task, it faces a number of challenges in noisy datasets and it is still considered an emerging research area~\cite{cano2013making, derczynski2015analysis}. Despite recent efforts, we still face limitations at \textit{identifying} entities and (consequently) correctly \textit{classifying} them. Current state-of-the-art NER systems typically have about 85-90\% accuracy on news text - such as articles (\textit{CoNLL03 shared task} data set) - but they still perform poorly (about 30-50\% accuracy) on short texts, which do not have implicit linguistic formalism (e.g. punctuation, spelling, spacing, formatting, unorthodox capitalisation, emoticons, abbreviations and hashtags)~\cite{ritter2011named,liu2012joint,Gattani2013,derczynski2015analysis}. Furthermore, the lack of external knowledge resources is an important gap in the process regardless of writing style~\cite{ratinov2009design}. To face these problems, research has been focusing on microblog-specific information extraction techniques~\cite{ritter2011named,van2013learning}.

In this paper, we propose a joint clustering architecture that aims at minimizing the current gap between world knowledge and knowledge available in open domain knowledge bases (e.g., Freebase) for NER systems, by extracting features from unstructured data sources. To this aim, we use images and text from the web as input data. Thus, instead of relying on encoded information and manually annotated resources (the major limitation in NER architectures) we focus on a multi-level approach for discovering named entities, combining text and image features with a final classifier based on a decision tree model. We follow an intuitive and simple idea: some types of images are more related to people (e.g. faces) whereas some others are more related to organisations (e.g. logos), for instance. This principle is applied similarly to the text retrieved from websites: keywords for search engines representing names and surnames of people will often return similarly related texts, for instance. Thus, we derive some indicators (detailed in \Cref{sec:finalclassifier} which are then used as input features in a final classifier.

To the best of our knowledge, this is the first report of a NER architecture which aims to provide \textit{a  priori} information based on clusters of images and text features. 
%The rest of the paper is structured as follows:~\Cref{sec:ner} introduces existing NER approaches.~\Cref{sec:architecture} presents our conceptual architecture, detailing its layers whereas~\Cref{sec:experiments} presents initial results over a noisy dataset. In ~\Cref{sec:discussion} we discuss relevant aspects of this approach. Finally,~\Cref{sec:conclusions} concludes with a brief summary of our work and future directions. 

\section{Related Work}\label{sec:ner}

Over the past few years, the problem of recognizing named entities in natural language texts has been addressed by several approaches and frameworks~\cite{nadeau2007survey,roberts2008combining}. Existing approaches basically adopt look-up strategies and use standard local features, such as \textit{part-of-speech tags}, \textit{previous and next words}, \textit{substrings},  \textit{shapes} and \textit{regex expressions}, for instance. The main drawback is the performance of those models with noisy data, such as Tweets. A major reason is that they rely heavily on hand-crafted features and domain-specific knowledge. In terms of architecture, NER algorithms may also be designed based on \textit{generative} (e.g., Naive Bayes) or \textit{discriminative} (e.g., MaxEnt) models. Furthermore, \textit{sequence} models (HMMs, CMM, MEMM and CRF) are a natural choice to design such systems. A more recent study proposed by Lample et al., 2016~\cite{lample2016neural} used neural architectures to solve this problem. Similarly in terms of architecture, Al-Rfou et al., 2015~\cite{al2015polyglot} had also proposed a model (without dependency) that learns distributed word representations (word embeddings) which encode semantic and syntactic features of words in each language. Chiu and Nichols, 2015~\cite{chiu2015named} proposed a neural network architecture that automatically detects word and character-level features using a hybrid bidirectional LSTM and CNN. Thus, these models work without resorting to any language-specific knowledge or resources such as \textit{gazetteers}. They, however, focused on \textit{newswire} to improve current state-of-the-art systems and not on the \textit{microblogs} context, in which they are naturally harder to outperform due to the aforementioned issues. According to Derczynski et al., 2015~\cite{derczynski2015analysis} some approaches have been proposed for Twitter, but they are mostly still in development and often not freely available.

\section{Conceptual Architecture}\label{sec:architecture}

\begin{figure*}[ht]
\centering
\includegraphics[width=1.0\textwidth]{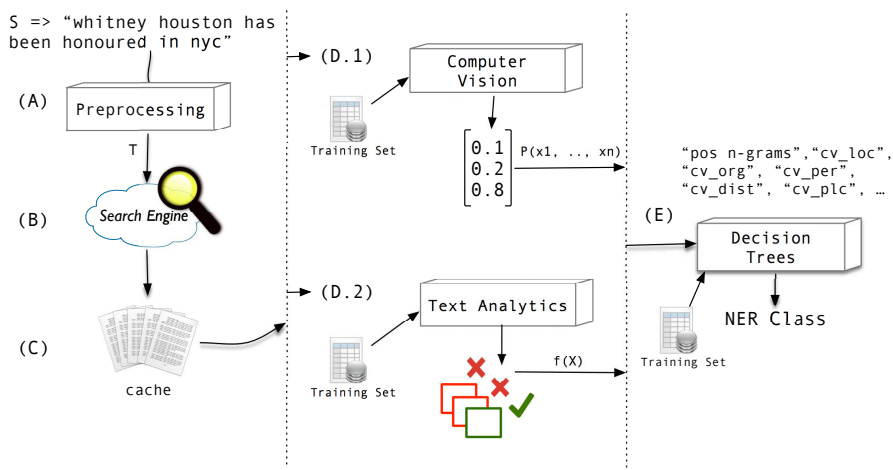}
\caption{Overview of the approach: combining computer vision and machine learning in a generic NER architecture}
\label{fig:architecture}
\end{figure*}

The main insight underlying this work is that we can produce a NER model which performs similarly to state-of-the-art approaches but without relying on any specific resource or encoded rule. To this aim, we propose a multi-level architecture which intends to produce biased indicators to a certain class (LOC, PER or ORG). These outcomes are then used as input features for our final classifier. We perform clustering on images and texts associated to a given \textit{term} $t$ existing in complete or partial sentences $S$ (e.g., ``new york'' or ``einstein''), leveraging the \textit{global context} obtained from the Web providing valuable insights apart from standard \textit{local features} and hand-coded information.~\Cref{fig:architecture} gives an overview of the proposed architecture.

In the first step \texttt{(A)}, we simply apply \textit{POS Tagging} and \textit{Shallow Parsing} to filter out tokens except for those\footnote{State-of-the-art POS tagging systems still do not have exceptional performance in short texts.} tagged as $propn$ or $nouns$ and their $compounds$ (\textit{local context}). Afterwards, we use the search engine (\texttt{B}) to query and cache (\texttt{C}) the top $N$ texts and images associated to each term $t \in T$, where $T$ is the set resulting of the pre-processing step (\texttt{A}) for each (partial or complete) sentence $S$. This resulting data (composed of excerpts of texts and images from web pages) is used to predict a possible class for a given term. These outcomes are then used in the first two levels (\texttt{D.1} and \texttt{D.2}) of our approach: the \textit{Computer Vision} and \textit{Text Analytics} components, respectively, which we introduce as follows:

\textbf{Computer Vision (CV): Detecting Objects:} \texttt{Function Description (D.1)}: given a set of images $\mathcal{I}$, the basic idea behind this component is to detect a specific object (denoted by a class $c$) in each image. Thus, we query the web for a given term $t$ and then extract the features from each image and try to detect a specific object (e.g., logos for \textit{ORG}) for the top $N$ images\footnote{We set $N$ = 10 in our experiments and used Microsoft Bing as the search engine.} retrieved as source candidates. The mapping between objects and NER classes is detailed in~\Cref{tab:tb_empirical}.
\begin{table}[ht]
	\centering
	\footnotesize
	\begin{tabular}{ll} 
		\toprule
	\textbf{NER} & \textbf{Images Candidates (number of trained models)} \\\midrule
	LOC & Building, Suburb, Street, City, Country, Mountain, Highway, Forest, Coast and Map (10)\\
	ORG & Company Logo (1)\\
	PER & Human Face (1)\\	 
		\bottomrule
	\end{tabular}
	\caption{NER classes and respective objects to be detected in a given image. For LOC we trained more models due to the diversity of the object candidates.}
	\label{tab:tb_empirical}
\end{table}
\texttt{Training (D.1)}: we used SIFT (Scale Invariant Feature Transform) features~\cite{790410} for extracting image descriptors and BoF (Bag of Features)~\cite{sivic2003video,philbin2007object} for clustering the histograms of extracted features. The clustering is possible by constructing a large vocabulary of many visual words and representing each image as a histogram of the frequency words that are in the image. We use \textit{k-means}~\cite{macqueen1967some} to cluster the set of descriptors to $k$ clusters. The resulting clusters are compact and separated by similar characteristics. An empirical analysis shows that some image groups are often related to certain named entities (NE) classes when using search engines, as described in ~\Cref{tab:tb_empirical}. For training purposes, we used the \textit{Scene 13} dataset~\cite{fei2005bayesian} to train our classifiers for \textit{location} (LOC), ``faces'' from \textit{Caltech 101 Object Categories}~\cite{fei2007learning} to train our \textit{person} (PER) and logos from \textit{METU dataset}~\cite{metulogo} for \textit{organisation} ORG object detection. These datasets produces the training data for our set of supervised classifiers (1 for ORG, 1 for PER and 10 for LOC). We trained our classifiers using Support Vector Machines~\cite{fletcher2009support} once they generalize reasonably enough for the task\footnote{\textit{scikit-learn}: svm.NuSVC(nu=0.5, kernel='rbf', gamma=0.1, probability=True).}.

\textbf{Text Analytics (TA): Text Classification} - \texttt{Function Description (D.2)}: analogously to \texttt{(D.1)}, we perform clustering to group texts together that are ``distributively'' similar. Thus, for each retrieved web page (title and excerpt of its content), we perform the classification based on the main NER classes. We extracted features using a classical sparse vectorizer (Term frequency-Inverse document frequency - TF-IDF. In experiments, we did not find a significant performance gain using \textit{HashingVectorizer}) - \texttt{Training (D.2)}: with this objective in mind, we trained classifiers that rely on a bag-of-words technique. We collected data using \textit{DBpedia} instances to create our training dataset ($N=15000$) and annotated each instance with the respective MUC classes, i.e. \textit{PER}, \textit{ORG} and \textit{LOC}. Listing \ref{lst:sparql} shows an example of a query to obtain documents of organizations (ORG class). Thereafter, we used this annotated dataset to train our model.

\begin{lstlisting}[captionpos=b, caption=SPARQL: an example of querying DBPedia to obtain LOC data for training, label=lst:sparql,
basicstyle=\ttfamily,frame=single]
SELECT ?location, ?abstract FROM <http://dbpedia.org>
WHERE {?location rdf:type dbo:Location .
       ?location dbo:abstract ?abstract .
FILTER (lang(?abstract) = 'en')} LIMIT 15000
\end{lstlisting}

\textbf{Final Classifier (E)}\label{sec:finalclassifier} - \texttt{Function Description (E)}: we use the outcomes of (\texttt{D.1} and \texttt{D.2}) as part of the input to our final model. The final set of indicators is defined as follows: let $W_{s}$ be a set of \textit{tokens} existing in a given sentence $s \in S$. We extract the POS tag (using Stanford POS Tagger) for each token $w$ and filter out any token classified other than \textit{PROP-NOUN} and existing \textit{compounds} as entity candidates ($t \in S\textprime$). The result is a simple structure:
\begin{equation}
M_{i} = \{j, t, ng_{pos}, C_{loc}, C_{per}, C_{org}, C_{dist}, C_{plc}, T_{loc}, T_{per}, T_{org}, T_{dist}\} 
\end{equation} 
where $i$ and $j$ represent the $i^{th}$ and $j^{th}$ position of $s \in S$ and $w \in W_{s}$, respectively. $ng_{pos}$ represents the n-gram\footnote{\textit{bigram}, in our experiments.} of POS tag. $C_{k}$ and $T_{k}$ ($k \in \{loc, per, org\}$) represent the total objects found by a classifier $\Phi$ for a given class $k$ ($\sum_{n=1}^{N} \Phi(k, img_{n})$)\footnote{$pos$ = +1, $neg$ = -1.} (where N is the total of retrieved images $\mathcal{I}$). $C_{dist}$ and $T_{dist}$ represent the distance between the two higher predictions ($\mathcal{P} = \{C_{k} \forall K\}$), i.e. $\max(\mathcal{P}) - \max(\mathcal{P}\textprime) | \mathcal{P}\textprime = \mathcal{P} - \{\max(\mathcal{P})\}$. Finally, $C_{plc}$ represents the sum of all predictions made by all $LOC$ classifiers $\mathcal{CL}$ ($\sum_{l=1}^{L} \sum_{n=1}^{N} \mathcal{CL}_{l}(loc,img_{n})$)\footnote{$pos$ = +1, $neg$ = 0.}. - \texttt{Training (E)}: the outcomes of \texttt{D.1} and \texttt{D.2} ($M$) are used as input features to our final classifier. We implemented a simple \textit{Decision Tree}\footnote{\textit{scikit-learn}: criterion='entropy', splitter='best'.} (non-parametric supervised learning method) algorithm for learning simple decision rules inferred from the data features (since it does not require any assumptions of linearity in the data and also works well with outliers, which are expected to be found more often in a noisy environment, such as the Web of Documents). 
 
\section{Experiments}
\label{sec:experiments}

In order to check the overall performance of the proposed technique, we ran our algorithm without any further rule or \textit{apriori} knowledge using a gold standard for NER in microblogs (Ritter dataset~\cite{ritter2011named}), achieving $0.59$ F1.~\Cref{tab:performance} details the performance measures per class.~\Cref{tab:relatedwork} presents current state-of-the-art results for the same dataset. The best model achieves $0.8$ F1-measure, but uses encoded rules. Models which are not rule-based, achieve $0.49$ and $0.56$. We argue that in combination with existing techniques (such as linguistic patterns), we can potentially achieve even better results.
\begin{table}[ht]
	\centering
	\footnotesize
	\begin{tabular}{lccc} 
		\toprule
		\textbf{NER Class} & \textbf{Precision} & \textbf{Recall} & \textbf{F-measure} \\\midrule
		Person (PER) & 0.86 & 0.53 & 0.66  \\
		Location (LOC) & 0.70 & 0.40 & 0.51 \\
		Organisation (ORG) & 0.90 & 0.46 & 0.61 \\
		None & 0.99 & 1.0 & 0.99 \\
		\midrule
		\rowcolor{lightgray}Average (PLO) & 0.82 & 0.46 & \textbf{0.59} \\
		\bottomrule
	\end{tabular}
	\caption{Performance measure for our approach in Ritter dataset: 4-fold cross validation}
	\label{tab:performance}
\end{table}
\begin{table}[ht]
	\centering
	\footnotesize
	\begin{tabular}{llccc} 
		\toprule
	\textbf{NER System} & \textbf{Description} & \textbf{Precision} & \textbf{Recall} & \textbf{F-measure} \\\midrule
	Ritter et al., 2011~\cite{ritter2011named} & LabeledLDA-Freebase  & 0.73 & 0.49 & 0.59 \\
	Bontcheva et al., 2013~\cite{bontcheva2013twitie}& Gazetteer/JAPE & 0.77 & 0.83 & 0.80  \\
	\rowcolor{lightgray} Bontcheva et al., 2013~\cite{bontcheva2013twitie}  & Stanford-twitter  & 0.54 & 0.45 & 0.49 \\
	\rowcolor{lightgray} Etter et al., 2013~\cite{etter2013nerit} & SVM-HMM & 0.65 & \textbf{0.49} & 0.54 \\
    \rowcolor{lightgray} \textit{our approach} & Cluster (images and texts) + DT & \textbf{0.82} & 0.46 & \textbf{0.59} \\
		\bottomrule
	\end{tabular}
\caption{Performance measures (PER, ORG and LOC classes) of state-of-the-art NER for short texts (Ritter dataset). Approaches which do not rely on hand-crafted rules and \textit{Gazetteers} are highlighted in gray. Etter et al., 2013 trained using 10 classes.}
\label{tab:relatedwork}
\end{table}
As an example, the sentence ``\textit{paris hilton was once the toast of the town}'' can show the potential of the proposed approach. The token ``\textit{paris}'' with a LOC bias (0.6) and ``\textit{hilton}'' (global brand of hotels and resorts) with indicators leading to LOC (0.7) or ORG (0.1, less likely though). Furthermore, ``\textit{town}'' being correctly biased to LOC (0.7). The algorithm also suggests that the \textit{compound} ``\textit{paris hilton}'' is more likely to be a PER instead (0.7) and updates (correctly) the previous predictions. As a downside in this example, the algorithm misclassified ``\textit{toast}'' as LOC. However, in this same example, Stanford NER annotates (mistakenly) only ``\textit{paris}'' as LOC. It is worth noting also the ability of the algorithm to take advantage of search engine capabilities. When searching for \textit{``miCRs0ft''}, the returned values strongly indicate a bias for ORG, as expected 
($C_{loc}$ = 0.2, $C_{org}$ = 0.8, $C_{per}$ = 0.0, $C_{dist}$ = 6, $C_{plc}$ = -56, $T_{loc}$ = 0.0, $T_{org}$ = 0.5, $T_{per}$ = 0.0, $T_{dist}$ = 5). More local organisations are also recognized correctly, such as ``\textit{kaufland}'' (German supermarket), which returns the following metadata: $C_{loc}$ = 0.2, $C_{org}$ = 0.4, $C_{per}$ = 0.0, $C_{dist}$ = 2, $C_{plc}$ = -50, $T_{loc}$ = 0.1, $T_{org}$ = 0.4, $T_{per}$ = 0.0, $T_{dist}$ = 3.

\section{Discussion}
\label{sec:discussion}

A disadvantage when using web search engines is that they are not open and free. This can be circumvented by indexing and searching on other large sources of information, such as Common Crawl and Flickr\footnote{\url{http://commoncrawl.org/} and \url{https://www.flickr.com/}}. However, maintaining a large source of images would be an issue, e.g. the Flickr dataset may not be comprehensive enough (i.e. tokens may not return results). This will be a subject of future work. Besides, an important step in the pre-processing is the classification of part-of-speech tags. In the Ritter dataset our current error propagation is 0.09 (107 tokens which should be classified as NOUN) using NLTK 3.0. Despite good performance (91\% accuracy), we plan to benchmark this component. In terms of processing time, the bottleneck of the current implementation is the time required to extract features from images, as expected. Currently we achieve a performance of 3\textasciitilde5 seconds per sentence and plan to also optimize this component. The major advantages of this approach are: 1) the fact that there are no hand-crafted rules encoded; 2) the ability to handle misspelled words (because the search engine alleviates that and returns relevant or related information) and incomplete sentences; 3) the generic design of its components, allowing multilingual processing with little effort (the only dependency is the POS tagger) and straightforward extension to support more NER classes (requiring a corpus of images and text associated to each desired NER class, which can be obtained from a Knowledge Base, such as DBpedia, and an image dataset, such as METU dataset). While initial results in a gold standard dataset showed the potential of the approach, we also plan to integrate these outcomes into a Sequence Labeling (SL) system, including neural architectures such as LSTM, which are more suitable for such tasks as NER or POS. We argue that this can potentially reduce the existing (significant) gap in NER performance on microblogs.

\section{Conclusions}
\label{sec:conclusions}

In this paper we presented a novel architecture for NER that expands the feature set space based on feature clustering of images and texts, focused on microblogs. Due to their terse nature, such noisy data often lack enough context, which poses a challenge to the correct identification of named entities. To address this issue we have presented and evaluated a novel approach using the Ritter dataset, showing consistent results over state-of-the-art models without using any external resource or encoded rule, achieving an average of 0.59 F1. The results slightly outperformed state-of-the-art models which do not rely on encoded rules (0.49 and 0.54 F1), suggesting the viability of using the produced metadata to also boost existing NER approaches. A further important contribution is the ability to handle single tokens and misspelled words successfully, which is of utmost importance in order to better understand short texts. Finally, the architecture of the approach and its indicators introduce potential to transparently support multilingual data, which is the subject of ongoing investigation. 

\section*{Acknowledgments}
This research was supported in part by an EU H2020 grant provided for the HOBBIT project (GA no. 688227) and CAPES Foundation (BEX 10179135).
%The acknowledgments should go immediately before the %references.  Do
%not number the acknowledgments section. Do not %include this section
%when submitting your paper for review.

\bibliographystyle{acm}
\bibliography{horus}

\end{document}